# Stability of spherical stellar systems I : Analytical results

Jérôme Perez [1,2] and Jean-Jacques Aly [1]

[1] *Service d'Astrophysique - CE Saclay - 91191 Gif sur Yvette cedex - France*
[2] *ETCA/CREA - 16 bis av. Prieur de la côte d'or - 94114 Arcueil cedex - France*

**ABSTRACT**
The so-called "symplectic method" is used for studying the linear stability of a self-gravitating collisionless stellar system, in which the particles are also submitted to an external potential. The system is steady and spherically symmetric, and its distribution function $f_0$ thus depends only on the energy $E$ and the squared angular momentum $L^2$ of a particle. Assuming that $\partial f_0/\partial E < 0$, it is first shown that stability holds with respect to all the spherical perturbations – a statement which turns out to be also valid for a rotating spherical system. Thus it is proven that the energy of an arbitrary aspherical perturbation associated to a "preserving generator" $\delta g_1$ [i.e., one satisfying $\partial f_0/\partial L^2 \{\delta g_1, L^2\} = 0$] is always positive if $\partial f_0/\partial L^2 \leq 0$ and the external mass density is a decreasing function of the distance $r$ to the center. This implies in particular (under the latter condition) the stability of an isotropic system with respect to all the perturbations.

Some new remarks on the relation between the symmetry of the system and the form of $f_0$ are also reported. It is argued in particular that a system with a distribution function of the form $f_0 = f_0(E, L^2)$ is necessarily spherically symmetric.

## 1 INTRODUCTION

The analytical study of the linear stability of a steady collisionless stellar system is a quite formidable task which has progressed with great difficulties since the pioneering work of Antonov ((Antonov 1962), (Antonov 1973)). The investigation methods which have been used up to now fall into two categories: The normal mode approach, and the energetic approach. The first one is presented in a systematic way in the most recent monograph by Palmer ((Palmer 1994), see also (Fridman & Polyachenko 1984) and references in these two books). It consists to derive from the linearized equations of motion (Vlasov-Poisson's system, VP hereafter) a dispersion relation for the eigenfrequencies of the system, and to try to extract from it as much information as possible on the nature of these numbers – an instability being present if at least one of them has a nonzero imaginary part, while stability holds if they are all real. The second method consists in constructing a quadratic functional $W$ over a set $\mathcal{Q}$ of admissible test functions satisfying some of the constraints actually fulfilled by any solution of VP. The system is found to be stable if $W$ keeps the same sign over $\mathcal{Q}$. Actually, there are several types of energy methods, which correspond to different choices for $W$ and $\mathcal{Q}$. The simplest one is the "thermodynamic method" ((Lynden-Bell & Sannit 1969), (Ipser 1974), (Ipser & Horwitz 1979)), in which only a very few constraints are taken into account. On the contrary, all the VP constraints are retained in the more elaborate "energy principle" ((Antonov 1962),(Kulsrud & Mark 1970), (Kandrup & Sygnet 1985)), as well as in the "symplectic method" ((Bartholomew 1971), (Kandrup 1990), (Kandrup 1991), (Perez 1995) and (Perez & Aly 1995), Paper I hereafter), the latter differing from the former by a restriction imposed to the perturbations, which are taken to be "symplectic" – i.e., they are generated by the infinitesimal canonical transforms acting on the steady state.

Our aim here is to reinvestigate the linear stability of a spherically symmetric system by fully working in the symplectic framework, with the hope of extending – or at least clarifying – some of the results which have been obtained thus far by the other energy methods. Our paper thus presents an interesting illustration of the general principles discussed in Paper I. At the same time, it also provides the theoretical background for the following paper of this series (Perez et al. 1995), in which the problem is tackled from a numerical point of view. It should be stressed from the very beginning that the method which is used here has not proven thus far to be able to reproduce all the results based on a normal mode analysis reported in (Palmer 1994). The latter, however, are sometimes based on approximations which appear quite difficult to justify rigorously, while ours are derived without the help of any approximation.

Our plan is as follows. In §2, we state precisely our assumptions – which include the possible presence of an external potential, generally not considered by previous authors –, and we develop some new considerations on the relations between the symmetry of the steady states we are interested in and the form of their distribution functions $f_0$. In §3, we recall the definition of a linear symplectic perturbation, the expression of its energy (which we cast into a particularly useful form), and the relations of the latter with the stability properties of the system. Thus we consider in details stability with respect to spherical (§4) and aspherical (§5) perturbations, respectively, assuming that $f_0$ is a decreasing function of the energy. Some extensions of our results to spherical rotating systems (Lynden-Bell 1960), and to systems with nonmonotonic $f_0$, are reported in §6.

## 2 EQUILIBRIUM

### 2.1 Notations and equations

We consider a system constituted of a large number of gravitating particles interacting together, and also submitted to an "external" potential $\Phi_e$. We denote as $\mathbf{x}$ and $\mathbf{v}$, respectively, the position and the velocity of a particle with respect to a Galilean frame $(O; \hat{\mathbf{x}}, \hat{\mathbf{y}}, \hat{\mathbf{z}})$, and as $\xi := (\mathbf{x}, \mathbf{v})$ the corresponding point in the phase space $\mathbf{R}^6$. Most often, we shall represent $\xi$ either by its standard spherical coordinates $(r, \theta, \phi, v_r, v_\theta, v_\phi)$ (with $\theta$ measured from the $z$-axis), or by its "canonical" coordinates $(r, \theta, \phi, p_r, p_\theta, p_\phi)$, with $p_r = v_r, p_\theta = rv_\theta, p_\phi = r\sin\theta v_\phi$ being the "conjugated" momenta of $(r, \theta, \phi)$ (e.g., (Landau & Lifchitz 1966)).

We assume that:

a) The external potential is spherically symmetric about $O$, i.e., $\Phi_e = \Phi_e(r)$. In most practical applications, $\Phi_e$ will represent the potential either of a central massive object or of a massive halo. For our theoretical purpose, however, we do not need to fix precisely its form, and we just consider it as being created by some matter distributed with the given smooth density $\rho_e(r)$. The total respective masses $M_e$ of that distribution and $M$ of the system, are finite.

b) The statistical state of the system is described at each time $t$ by a one-particle distribution function $\mathcal{F}(\xi, t, m)$, with $\mathcal{F}(\xi, t, m) \, d\xi dm$ representing the number of particles of mass between $m$ and $m + dm$ contained in the phase space volume $d\xi$ around $\xi$.

c) Collisions between the particles are negligible. Then it is sufficient to consider the mass-averaged distribution function

$$f(\xi, t) := \int m \mathcal{F}(\xi, t, m) \, dm \, . \tag{1}$$

$f$ solves the VP system

$$\frac{\partial f}{\partial t} + \mathbf{v}.\nabla_x f - \nabla_x \Psi.\nabla_v f = \frac{\partial f}{\partial t} + \{f, h\} = 0 \, , \tag{2}$$

$$\nabla_x^2 \Psi = 4\pi G(\rho + \rho_e) := 4\pi G \left( \int f \, d\mathbf{v} + \rho_e \right) \, , \tag{3}$$

$$\Psi =_{r \to \infty} O(r^{-1}) \quad \text{and} \quad \lim_{(r,v) \to \infty} f(\mathbf{x}, \mathbf{v}) = 0 \, , \tag{4}$$

with the decrease of $f$ at infinity being supposed to be sufficiently fast. Here, $\Psi := \Phi + \Phi_e$ is the total gravitational potential created by the particles and the external masses,

$$h(\mathbf{x}, \mathbf{v}, t) := \frac{v^2}{2} + \Psi(\mathbf{x}, t) \tag{5}$$

is the one-particle Hamiltonian, and $\{.,.\}$ denotes the Poisson bracket, defined by

$$\{f_1, f_2\} := \nabla_x f_1.\nabla_v f_2 - \nabla_v f_1.\nabla_x f_2 \, . \tag{6}$$

Note that there is no explicit dependance on the mass of the particles appearing in Eqs. (2)-(3).

Instead of (2)-(3), we may use only Eq. (2) in which we have substituted for $\Phi$ the unique solution to Eqs. (3)-(4). The latter can be written in integral form as $\Phi = K[f]$, with the operator $K$ being quite generally defined by

$$K[f](\mathbf{x}) := -G \int \frac{f(\xi')}{|\mathbf{x} - \mathbf{x}'|} \, d\xi' . \tag{7}$$

### 2.2 Spherically symmetric steady states

We are interested here in a steady state which is spherically symmetric with respect to $O$, and then, in particular, occupies a sphere of radius $R$ ($\leq \infty$) in the physical space. Hence $f_0$ – assumed from now on to be a sufficiently smooth function – and the associated potential $\Phi_0 = K[f_0]$ are left unchanged by any rotation of $\mathbf{R}^3$ of center $O$, which implies that they can depend only on the three geometrical invariants $r$, $v$ and $\mathbf{x}.\mathbf{v} = rv_r$ – i.e., they are necessarily of the form $f_0 = f_0(r, v, rv_r)$ and $\Phi_0 = \Phi_0(r)$, respectively.

Let us introduce the two quantities (we set $\mathbf{v}_t := \mathbf{v} - v_r \hat{\mathbf{r}}$)

$$E := h_0 := \frac{v^2}{2} + \Psi_0(r) := \frac{v_r^2}{2} + \frac{v_t^2}{2} + \Phi_0(r) + \Phi_e(r) \tag{8}$$

and

$$L^2 := r^2 v_t^2 := r^2(v_\theta^2 + v_\phi^2) = p_\theta^2 + \frac{p_\phi^2}{\sin^2\theta}, \tag{9}$$

which represent the energy of a particle in the potential $\Psi_0$ and the squarred modulus of its angular momentum with respect to $O$. Clearly, we can express $v$ in terms of $r$, $v_r$ and $L^2$, and thus write

$$f_0 = f_0(r, v_r, L^2). \tag{10}$$

Injecting this expression into the time-independant version of Eq. (2), we obtain

$$v_r \frac{\partial f_0}{\partial r} + \left(\frac{L^2}{r^2} - \Psi_0'\right) \frac{\partial f_0}{\partial v_r} = 0. \tag{11}$$

From the theory of first order partial differential equations, we know that the general solution of Eq. (11) is a function of two independant integrals of its characteristic system

$$\frac{\mathrm{d}r}{v_r} = \frac{\mathrm{d}v_r}{L^2/r^2 - \Psi_0'}. \tag{12}$$

Two such integrals are obvious here: $L^2$ (as $\mathrm{d}L^2$ does not appear), and $E(r, v_r, L^2)$. Thus we can conclude at once that

$$f_0 = f_0(E, L^2). \tag{13}$$

The particular case $f_0 = f_0(E)$ corresponds to an isotropic equilibrium.

The general form (13) of $f_0$ is well known, but it is usually derived (e.g., (Dejonghe 1986), (Binney & Tremaine 1987), (Palmer 1994)) by appealing to the so-called strong Jeans theorem, which deals with the isolating integrals of motion. As this theorem is far from being easy both to derive and to apply, we felt that it was a useful task to provide the reader with a completely straigthforward proof. In particular, the proof presented here shows most clearly the basic reasons why $f_0$ depends on only two variables: Symmetry requires $f_0$ to depends on only three variables, and the steady VP reduces this number by one unit.

Evidently, the potential $\Psi_0$ appearing in $E$ is a solution of

$$\nabla_x^2 \Psi_0 = \frac{1}{r^2} \frac{\mathrm{d}}{\mathrm{d}r}\left(r^2 \frac{\mathrm{d}\Psi_0}{\mathrm{d}r}\right) = 4\pi G(\rho_0 + \rho_e) := 4\pi G \left(\int f_0(E, L^2) \, \mathrm{d}\mathbf{v} + \rho_e\right), \tag{14}$$

$$\Psi_0 =_{r\to\infty} O(r^{-1}). \tag{15}$$

An immediate consequence (which will prove quite important hereafter) of these equations is that $\Psi_0(r)$ is a strictly increasing negative function (as $\rho \geq 0$ and $\rho_e \geq 0$). $\Phi_0$ and $\Phi_e$ (if $\rho_e \not\equiv 0$) have obviously the same property.

The characteristics of the steady state VP describe the motion of the individual particles in the (self-consistent) potential $\Psi_0$. In particular, if we substitute $\mathrm{d}r/\mathrm{d}t$ for $v_r$, Eq. (12) and its integral $E$ determine the radial motion of a particle of energy $E$ and angular momentum modulus $L$, which can be described as a one-dimensional motion in the effective potential $\Psi_{eff} := \Psi_0 + L^2/2r^2$. When $r$ increases from 0 to $+\infty$, the latter decreases from $+\infty$ to a negative minimum, reached for $r_0(L^2)$, and then increases to 0 $^\star$. As the system is confined inside the sphere of radius $R$ and $v_r^2 \geq 0$, we have necessarily

$$E^-(L^2) := \Psi_{eff}[r_0(L^2)] = \Psi_0[r_0(L^2)] + \frac{L^2}{2[r_0(L^2)]^2} \leq E \leq \Psi_0(R) + \frac{L^2}{2R^2} =: E^+(L^2) \leq 0 \tag{16}$$

[with $E^+(L^2) = 0$ if $R = \infty$; for an isotropic state, we can even take $E^+(L^2) = \Psi_0(R)$, as $v_r = 0$ implies $\mathbf{v} = 0$] – the particle thus bouncing between the two solutions $r^-(E, L^2)$ and $r^+(E, L^2)$ $[r^- \leq r_0 \leq r^+ \leq R]$ to the equation $\Psi_{eff}(r) = E$. Eq. (16) imposes $L^2$ to be bounded from above by the smallest solution $L^2_{max}$ of the equation $E^-(L^2) = E^+(L^2)$, or equivalently of $r_0(L^2_{max}) = R$ [the existence of that solution results immediately from the monotonic increase of $r_0(L^2)$ $^\star$]. Then we have eventually reached the conclusion that $f_0(E, L^2)$ must vanish outside the energy range $[E^-(L^2), E^+(L^2)]$, where $0 \leq L^2 \leq L^2_{max}$.

Hereafter, we shall denote as $\Omega_0$ the region of the phase space where $f_0 > 0$, and set

$$f_{0E} := \frac{\partial f_0}{\partial E} \quad \text{and} \quad f_{0L^2} := \frac{\partial f_0}{\partial L^2}. \tag{17}$$

---

$^\star$ This last statement is quoted very often in the literature, but without proof. To establish it, we remark that $\Psi'_{eff} = 0$ implies $G(M_r + M_{er}) = L^2/r$, where $(M_r + M_{er})$ denotes the total mass in a sphere of radius $r$, and we have used the integral of Eq. (14) (Newton's theorem); as the LHS of that equation is a nondecreasing function of $r$, while its RHS is a decreasing one, it has one and only one solution $r_0(L^2)$. The conclusion thus follows by noting the obvious behaviour of $\Psi_{eff}$ when $r \to 0$ and $r \to \infty$. It is also interesting to note that $r_0(L^2)$ increases monotonically with $L^2$ ($\mathrm{d}r_0/\mathrm{d}L^2 > 0$) from 0 to $\infty$, with $r_0 \propto L^{1/2}$ when $L^2 \to 0$, and $r_0 \propto L^2$ when $L^2 \to \infty$.

Unless otherwise specified, it will be assumed that

$$f_{0E} < 0 \tag{18}$$

in $\Omega_0$.

### 2.3 Symmetry of $f_0$: Converse theorems

In the previous subsection, we have shown that a steady spherically symmetric system has an $f_0$ of the form (13). It is then natural to address the converse question: *If a steady state is characterized by a distribution function of the form $f_0(E, L^2)$, is it necessarily spherically symmetric?*

Let us then assume that we have some steady state described by a sufficiently smooth distribution function $f_0(E, L^2)$ [condition (18) is not imposed here]. Hence, $f_0$ does satisfy

$$\{f_0, E\} = f_{0E}\{E, E\} + f_{0L^2}\{L^2, E\} = -2 f_{0L^2} \mathbf{v}_t . \nabla \Psi_0(\mathbf{x}) = 0 , \tag{19}$$

while $u := -\Psi_0(\mathbf{x})$ is a positive solution (with no a priori prescribed symmetry) to the nonlinear boundary value problem

$$-\nabla_x^2 u(\mathbf{x}) = 4\pi G \left\{ \int f_0 \left( \frac{v^2}{2} - u(\mathbf{x}), r^2 v_t^2 \right) d\mathbf{v} + \rho_e(r) \right\} =: g[u(\mathbf{x}), r] , \tag{20}$$

$$u =_{r \to \infty} O(r^{-1}) . \tag{21}$$

The function $g$ defined in Eq. (20) is supposed to be sufficiently regular. Note that we have distinguished in the dependance of $g$ on $\mathbf{x}$ a part which arises from the dependance of $f_0$ on $E$, and a part which is related to both the dependances of $f_0$ on $L^2$ and of $\rho_e$ on $r$ (this separation being clearly possible without any ambiguity).

Consider first the case where $f_{0L^2} \not\equiv 0$. Then Eq. (19) implies at once that $\Psi_0(\mathbf{x}) = \Psi_0(r)$ – i.e., spherical symmetry holds indeed – if, whenever $\rho_0(\mathbf{x}) > 0$, there is a $\mathbf{v}$ such that $f_{0L^2}(\mathbf{x}, \mathbf{v}) \neq 0$ [this situation is certainly met if $f_{0L^2} \neq 0$ in $\Omega_0$ – except maybe on a negligible set of points –, which happens, for instance, if $f_0$ is of one of the standard forms $f_0(E \pm L^2/2r_a^2)$, $f_1(E) L^{2\alpha}$, … (Binney & Tremaine 1987)]. If the condition above is not satisfied, Eq. (19) only implies that $\Psi_0(\mathbf{x}) = \Psi_0(r)$ inside some spherical shells, but the general conclusion can still be reached by applying a continuation argument to Eq. (20) (note that both $\Psi_0$ and its normal derivative $\partial_r \Psi_0$ take constant values on the boundary of the shells).

The situation appears to be more intricate when $f_0$ does not depend explicitly on $L^2$ ($f_{0L^2} \equiv 0$). But fortunately, we may then use some recent mathematical results which apply to the solutions to Eqs. (20)-(21) under some technical assumptions (Gidas et al. 1981) (the interested reader is referred to that paper for complete statements and proofs of the theorems we shall use). Let us first assume that $\rho_e \equiv 0$ and $g(u) = O(u^\alpha)$ near $u = 0$, whith $\alpha > 4$ [the last condition is satisfied in most cases of practical interest; for instance, it holds true when $f_0(E) = 0$ for $E^+ \leq E$, with the constant $E^+ < 0$ – in which case the system is contained in a bounded volume –, and for Plummer's model, where $g(u) \propto u^5$ (Binney & Tremaine 1987)]. Then we are in the conditions of Theorem 1 of (Gidas et al. 1981), and we can assert at once that $u$, and then $\Psi_0$, is necessarily spherically symmetric about the origin $O$ (possibly after a translation). The fact that $du/dr = -d\Psi_0/dr < 0$ for $r > 0$, also asserted by the theorem, results more directly here from $g$ being nonnegative (see §2.2). If $\rho_e \not\equiv 0$, symmetry around $O$ is obtained as a consequence of theorem 1" of (Gidas et al. 1981) (after some slight changes made possible by the particular form of our $g$); we just need to assume that $\rho'_e \leq 0$, $\rho_0 \rho_e \neq 0$ at some point, and $g(u, r) \leq c u^\alpha$ near $u = 0$, with $c > 0$ and $\alpha > 4$.

We have thus reached eventually the conclusion that, under reasonable assumptions, a state characterized by $f_0(E, L^2)$ is spherically symmetric indeed.

## 3 SYMPLECTIC ENERGY FUNCTIONAL AND LINEAR STABILITY

### 3.1 Symplectic perturbations

Let us now recall some of the basic facts about the symplectic approach to the stability of a steady state, referring the readers to (Bartholomew 1971), (Kandrup 1990), (Kandrup 1991) and Paper I for details. A linear "symplectic" perturbation is defined to be a first order change of the distribution function which is of the form

$$f_1 = -\{f_0, g_1\} , \tag{22}$$

for some arbitrary regular phase-space function $g_1$ – the so-called "generator" –, that it is convenient to consider formally as being complex valued. Clearly, a symplectic perturbation admits of an infinity of generators, two of them differing by some function $\tilde{g}$ commuting with $f_0$ ($\{f_0, \tilde{g}\} = 0$). We shall say that $g_1$ is nontrivial if $f_1 \not\equiv 0$.

To any $g_1$ are also associated a first order variation of the potential,

$$\Phi_1 = K[f_1] = K[-\{f_0, g_1\}], \tag{23}$$

and a variation of the total energy which turns out to be second order in $g_1$, being given by

$$H^{(2)}[g_1] = -\frac{1}{2} \int \{f_0, g_1\}\{E, g_1^*\} \, d\xi - \frac{1}{8\pi G} \int |\nabla \Phi_1|^2 \, d\mathbf{x} \tag{24}$$

("*" denotes complex conjugation). It is worth noticing that:

a) $H^{(2)}[g_1]$ is a conserved quantity if $g_1$ is taken to be a solution of the linearized equation of motion

$$\frac{\partial g_1}{\partial t} = \mathcal{L}[g_1] := \{E, g_1\} + K[\{f_0, g_1\}] \tag{25}$$

[in which case $f_1 = -\{f_0, g_1\}$ solves the linearized version of VP].

b) If the system is translated as a whole by an infinitesimal vector $\mathbf{a}$ ($g_1 = \mathbf{a}.\mathbf{v}$ and $f_1 = -\mathbf{a}.\nabla f_0$), then $H^{(2)}[g_1] = 0$ in the absence of external potential (this is intuitively obvious, and easy to check formally). This type of "neutral" displacements being of no interest, we shall eliminate them by imposing the admissible generators, whose set will be denoted by $\mathcal{G}$, to satisfy

$$\int f_1 \mathbf{x} \, d\xi = \int g_1 \nabla_v f_0 \, d\xi = 0 \tag{26}$$

in the case where $\Phi_e \equiv 0$ – this condition impeding the center of mass of the system to be moved by the perturbation.

## 3.2 Linear stability criteria

The stability properties of the steady state turn out to be strongly related to the values taken by $H^{(2)}$ on $\mathcal{G}$, and it is always an important first step when considering stability to classify as precisely as possible the generators $g_1$ according to the sign of that quantity. In fact, this study may even provide in some cases useful <u>sufficient</u> conditions of stability ((Bartholomew 1971), (Kandrup 1990), (Kandrup 1991) and Paper I):

(a) If

$$H^{(2)}[g_1] > 0 \ (\text{or} < 0) \tag{27}$$

for any nontrivial $g_1$ of $\mathcal{G}$, then the system is linearly stable with respect to any perturbation. Unfortunately, this condition may be possibly satisfied only if $f_0 = f_0(E)$ and $f_{0E} \leq 0$ (actually, it may be fulfilled for a more general form of $f_0$, corresponding to a uniformly rotating system). Indeed, if $f_0$ depends explicitly on some other integral, there does always exist both positive and negative energy perturbations [Paper I. The main step in proving this result is the construction of two phase space functions $\alpha$ and $\beta$ and of a number $\epsilon$ such that the first term in the RHS of Eq. (24) be positive (resp., negative) for the generator $g_1 := \alpha e^{i\beta/\epsilon}$].

(b) Suppose that we cannot prove Eq. (27) for all the admissible $g_1$, but that we can identify in $\mathcal{G}$ a linear subspace $\mathcal{G}'$ that is "closed" under evolution – which means that, if an arbitrary $g_1 \in \mathcal{G}'$ is taken as an initial condition for an evolution governed by Eq. (25), then the solution $g_1(t)$ stays in $\mathcal{G}'$ forever]. Then the system is stable with respect to the corresponding restricted class of perturbations if Eq. (27) holds true for all the nontrivial elements of $\mathcal{G}'$.

## 3.3 Decomposition of $H^{(2)}$

For our spherically symmetric system, it will prove convenient to transform the expression for $H^{(2)}[g_1]$ into a different form. For that, we first introduce the following notations associated to the arbitrary phase function $g$ and the arbitrary rotation $\mathcal{R}$ of $\mathbf{R}^3$ about $O$:

$$g_\mathcal{R}(\xi) = g(\mathcal{R}\mathbf{x}, \mathcal{R}\mathbf{v}), \tag{28}$$

$$\overline{g}(\xi) = \frac{1}{4\pi} \int g_\mathcal{R}(\xi) \, d\mathcal{R}, \tag{29}$$

$$\delta g = g - \overline{g}. \tag{30}$$

Clearly, the average $\overline{g}$ is spherically symmetric, i.e., $\overline{g}(\mathcal{R}\xi) = \overline{g}(\xi)$ for any $\mathcal{R}$, and $\overline{g} = \overline{g}(r, v, rv_r)$. We have also the useful simple relations

$$\overline{\delta g} = \frac{1}{4\pi} \int (\delta g)_\mathcal{R} \, d\mathcal{R} = 0, \tag{31}$$

and, for any $g$ and $g'$

$$\{g_\mathcal{R}, g'_\mathcal{R}\}(\xi) = \{g, g'\}(\mathcal{R}\xi). \tag{32}$$

For a perturbation generated by $g_1$, we can thus write

$$g_1 = \overline{g_1} + \delta g_1, \tag{33}$$

$$f_1 = \overline{f_1} + \delta f_1, \tag{34}$$

$$\Phi_1 = \overline{\Phi_1} + \delta\Phi_1 \, . \tag{35}$$

Using the symmetry of $f_0$ and $E$, the rotational invariance of the Laplacian and the properties of the averaging process quoted above, we obtain immediately

$$\overline{f_1} = -\{f_0, \overline{g_1}\} = -f_{0E}\{E, \overline{g_1}\} \quad \text{and} \quad \delta f_1 = -\{f_0, \delta g_1\} \, , \tag{36}$$

$$\overline{\Phi_1} = K[-\{f_0, \overline{g_1}\}] \quad \text{and} \quad \delta\Phi_1 = K[-\{f_0, \delta g_1\}] \, , \tag{37}$$

and after some straightforward algebra,

$$H^{(2)}[g_1] = H^{(2)}[\overline{g_1}] + H^{(2)}[\delta g_1] \, , \tag{38}$$

with of course

$$H^{(2)}[\overline{g_1}] = \frac{1}{2} \int (-f_{0E}) |\{E, \overline{g_1}\}|^2 \, d\xi - \frac{1}{8\pi G} \int |\nabla_x \overline{\Phi_1}|^2 \, d\mathbf{x} \, , \tag{39}$$

$$H^{(2)}[\delta g_1] = -\frac{1}{2} \int \{f_0, \delta g_1\}\{E, \delta g_1^*\} \, d\xi - \frac{1}{8\pi G} \int |\nabla_x \delta\Phi_1|^2 \, d\mathbf{x} \, . \tag{40}$$

We now consider in turn each of the two terms appearing in the RHS of Eq. (38).

## 4 STABILITY WITH RESPECT TO SPHERICAL PERTURBATIONS

### 4.1 A further decomposition

For studying the term $H^{(2)}[\overline{g_1}]$, it is useful to effect one further decomposition by proceeding as follows. For any average $\overline{g}$, we denote as $\overline{g}^{+/-}$, respectively, its symmetric and antisymmetric parts (with respect to the transform $v_r \to -v_r$), defined by

$$\overline{g}(r, v, rv_r) = \frac{1}{2}[\overline{g}(r, v, rv_r) + \overline{g}(r, v, -rv_r)] + \frac{1}{2}[\overline{g}(r, v, rv_r) - \overline{g}(r, v, -rv_r)] =: \overline{g}^+(r, v, rv_r) + \overline{g}^-(r, v, rv_r) \, . \tag{41}$$

With

$$\overline{g_1} = \overline{g_1}^+ + \overline{g_1}^- \, , \tag{42}$$

$$\overline{f_1} = \overline{f_1}^+ + \overline{f_1}^- \, , \tag{43}$$

we thus have

$$\overline{f_1}^\pm = -\{f_0, \overline{g_1}^\mp\} \, , \tag{44}$$

$$\overline{\Phi_1} = K[-\{f_0, \overline{g_1}^-\}] \, , \tag{45}$$

and we get after some simple algebra,

$$H^{(2)}[\overline{g_1}] = H^{(2)}[\overline{g_1}^+] + H^{(2)}[\overline{g_1}^-] \, , \tag{46}$$

with

$$H^{(2)}[\overline{g_1}^+] = \frac{1}{2} \int (-f_{0E}) |\{E, \overline{g_1}^+\}|^2 \, d\xi \, , \tag{47}$$

$$H^{(2)}[\overline{g_1}^-] = \frac{1}{2} \int (-f_{0E}) |\{E, \overline{g_1}^-\}|^2 \, d\xi - \frac{1}{8\pi G} \int |\nabla_x \overline{\Phi_1}|^2 \, d\mathbf{x} \, . \tag{48}$$

The first part $H^{(2)}[\overline{g_1}^+]$ of $H^{(2)}[\overline{g_1}]$ is clearly nonnegative, and thus we just need to consider in details the second part $H^{(2)}[\overline{g_1}^-]$.

### 4.2 The sign of $H^{(2)}[\overline{g_1}^-]$

To study the sign of $H^{(2)}[\overline{g_1}^-]$, we closely follow the method of (Sygnet et al. 1984), which originates in the work of (Gillon et al. 1976). However, in addition to the facts that we have an external potential $\Phi_e$ and that we do not use $(r, L^2, E)$ as independant variables, there are some essential differences between their calculations and ours, that we shall discuss below.

First, we transform the expression of $H^{(2)}[\overline{g_1}^-]$ into a more convenient form. For that, we integrate once Poisson's equation

$$\nabla_x^2 \overline{\Phi_1} = \frac{1}{r^2} \frac{d}{dr} r^2 \frac{d\overline{\Phi_1}}{dr} = -4\pi G \int \{f_0, \overline{g_1}^-\} \, d\mathbf{v} = 4\pi G \frac{1}{r^2} \frac{d}{dr}\left(r^2 \int f_{0E} v_r \overline{g_1}^- \, d\mathbf{v}\right) \, , \tag{49}$$

(where we have effected two integrations by part in the second member to get the RHS), which gives

$$\nabla \overline{\Phi_1} = \frac{\mathrm{d}\overline{\Phi_1}}{\mathrm{d}r}\hat{\mathbf{r}} = \left(4\pi G \int f_{0E} v_r \overline{g_1}^- \, \mathrm{d}\mathbf{v}\right) \hat{\mathbf{r}}. \tag{50}$$

Injecting this result into the second term of the RHS of Eq. (39) and using Schwartz's inequality, we obtain

$$\begin{aligned}
\frac{1}{8\pi G} \int |\nabla_x \overline{\Phi_1}|^2 \, \mathrm{d}\mathbf{x} &= 2\pi G \int \left| \int f_{0E} v_r \overline{g_1}^- \, \mathrm{d}\mathbf{v} \right|^2 \mathrm{d}\mathbf{x} \\
&\leq 2\pi G \int \left( \int (-f_{0E}) v_r^2 \, \mathrm{d}\mathbf{v} \right) \left( \int (-f_{0E}) |\overline{g_1}^-|^2 \, \mathrm{d}\mathbf{v} \right) \mathrm{d}\mathbf{x}.
\end{aligned} \tag{51}$$

Noticing that

$$\rho_0 = \int f_0 \, \mathrm{d}\mathbf{v} = \int (-f_{0E}) v_r^2 \, \mathrm{d}\mathbf{v} \tag{52}$$

(just remark that $\mathrm{d}\mathbf{v} = 2\pi v_t \, \mathrm{d}v_r \mathrm{d}v_t$ and integrate by part with respect to $v_r$), we thus obtain

$$H^{(2)}\{\overline{g_1}\} \geq \frac{1}{2} \int (-f_{0E}) \left( |\{E, \overline{g_1}^-\}|^2 - 4\pi G \rho_0 |\overline{g_1}^-|^2 \right) \, \mathrm{d}\xi. \tag{53}$$

We now make the change of variable

$$\overline{g_1}^- =: r v_r \mu, \tag{54}$$

whith the new function $\mu$ being still regular for $rv_r = 0$ as $\overline{g_1}^-(r,v,rv_r = 0) = 0$ owing to the antisymmetry of $\overline{g_1}^-$. Using the derivative property of the Poisson bracket, and the fact that the integral over the phase space of a Poisson bracket vanishes, we can rewrite Eq. (53) in the form

$$\begin{aligned}
H^{(2)}[\overline{g_1}^-] &\geq \frac{1}{2} \int (-f_{0E}) \left[ (rv_r)^2 |\{E,\mu\}|^2 + rv_r \{E, rv_r\}\{E, |\mu|^2\} + |\mu|^2 \{E, rv_r\}^2 - 4\pi G\rho_0 (rv_r|\mu|)^2 \right] \mathrm{d}\xi \\
&= \frac{1}{2} \int (-f_{0E}) \left[ (rv_r)^2 |\{E,\mu\}|^2 + \{E, |\mu|^2 rv_r \{E, rv_r\}\} - |\mu|^2 rv_r \{E, \{E, rv_r\}\} \right. \\
&\quad \left. - |\mu|^2 \{E, rv_r\}^2 + |\mu|^2 \{E, rv_r\}^2 - 4\pi G\rho_0 |rv_r \mu|^2 \right] \mathrm{d}\xi \\
&= \frac{1}{2} \int (-f_{0E}) \left[ (rv_r)^2 |\{E,\mu\}|^2 - |\mu|^2 \left( rv_r\{E, \{E, rv_r\}\} + 4\pi G\rho_0 (rv_r)^2 \right) \right] \mathrm{d}\xi.
\end{aligned} \tag{55}$$

By a straightforward calculation, we obtain

$$\{E, \{E, rv_r\}\} = -rv_r \left( \frac{\mathrm{d}^2 \Psi_0}{\mathrm{d}r^2} + \frac{3}{r}\frac{\mathrm{d}\Psi_0}{\mathrm{d}r} \right) = -rv_r \left( 4\pi G(\rho_0 + \rho_e) + \frac{1}{r}\frac{\mathrm{d}\Psi_0}{\mathrm{d}r} \right), \tag{56}$$

whence, injecting this expression into Eq. (55),

$$H^{(2)}[\overline{g_1}^-] \geq \frac{1}{2} \int (-f_{0E})(rv_r)^2 \left[ |\{E,\mu\}|^2 + |\mu|^2 \left( 4\pi G\rho_e + \frac{1}{r}\frac{\mathrm{d}\Psi_0}{\mathrm{d}r} \right) \right] \mathrm{d}\xi \geq 0, \tag{57}$$

the last inequality sign resulting from the negativeness of $f_{0E}$ and the positiveness of $\rho_e$ and $\mathrm{d}\Psi_0/\mathrm{d}r$.

Let us now assume that $H^{(2)}[\overline{g_1}] = 0$. Then $H^{(2)}[\overline{g_1}^+] = 0$, whence $\{f_0, \overline{g_1}^+\} = 0$; and $H^{(2)}[\overline{g_1}^-] = 0$, whence $f_{0E}\overline{g_1}^- = 0$, and $\{f_0, \overline{g_1}^-\} = 0$. Therefore $\overline{f_1} = 0$, and we can eventually conclude that $H^{(2)}[\overline{g_1}] > 0$ for any $g_1$ such that $\overline{f_1} = -\{f_0, \overline{g_1}\} \not\equiv 0$.

We can summarize the results of this section in the form of a theorem:

*Theorem 1:* *If the steady state satisfies condition (18), then $H^{(2)}[\overline{g_1}] > 0$ for any $g_1$ with a nontrivial average $\overline{g_1}$.*

### 4.3 Consequences for stability

It is quite obvious that the set $\mathcal{G}'$ of the generators underlying the spherically symmetric perturbations is "closed with respect to evolution" in the sense of §3.2 – i.e., the solution $g_1(t)$ to Eq. (25) generates a spherically symmetric perturbations if the initial condition $g_1(0)$ does. Then, by the general statement (b) of §3.2, we can conclude at once that *any spherical steady state satisfying (18) is linearly stable with respect to all the spherical perturbations.*

### 4.4 Comparison with previous works

Of course, this result has been reported many times in the literature for the case where $\Phi_e \equiv 0$, and it is then worth explaining in a few words the differences between our proof and previous ones. In fact, in the framework of the energy methods, the game consists in any case to show the positiveness of some quadratic functional $W[q]$ over a set $\mathcal{Q}$ of admissible functions $q$.

a) In a first class of methods ((Antonov 1962), (Kulsrud & Mark 1970), (Kandrup & Sygnet 1985)), $\mathcal{Q}_1$ is taken to be the set of all the functions which vanish outside $\Omega_0$ and are antisymmetric in the change $v_r \to -v_r$, to which belongs the antisymmetric part $\overline{f_1}^-$ of any solution to the linearized VP. The relevant functional $W_1$ may be derived in a standard way (e.g., (Laval et al. 1965)) from the equation of motion (second order in time!)

$$\frac{1}{-f_{0E}} \frac{\partial^2 \overline{f_1}^-}{\partial t^2} = \mathcal{M} \overline{f_1}^- \tag{58}$$

satisfied by $\overline{f_1}^-$, the essential point being here the self-adjoint character (with respect to the usual hermitian product) of the operator $\mathcal{M}$ (Kandrup & Sygnet 1985). It turns out that

$$W_1[\overline{f_1}^-] = H^{(2)}[\overline{f_1}^-]. \tag{59}$$

It appears that the "energy principle" method and the symplectic one lead to formally identical functionals in the situation we are considering here. They can thus be treated by the similar technics, and the same results are obtained. However, the two methods are conceptually very different ((Bartholomew 1971), (Kandrup 1990), (Kandrup 1991), Paper I). In the former, the derived "energy functional" is closely related to the peculiar properties of the second order equation one starts with [the existence of $W_1$ results from the self-adjointness of $\mathcal{M}$, which is just a consequence of the strong symmetries of the system], and it has no obvious physical meaning. The latter, on the contrary, is more transparent. It only deals with a first order equation – the one satisfied by the generators $g_1$ of the symplectic perturbations (which form a restricted class of perturbations) –, and $H^{(2)}[\overline{g_1}]$ is easily seen to be actually the energy of the perturbation – an interpretation which keeps its validity when one considers more general systems.

b) In a second class of methods ((Lynden-Bell & Sannit 1969), (Ipser 1974), (Ipser & Horwitz 1979)), the relevant functional $W_2$ is the second variation of $E[f] + H_C[f]$. Here, $E[f]$ is the energy of $f$ and

$$H_C[f] := -\int C[f, L^2] \, \mathrm{d}\xi := \int \left( \int_0^f E(f_0, L^2) \, \mathrm{d}f_0 \right) \mathrm{d}\xi , \tag{60}$$

with $E(f_0, L^2)$ being the "inverse" of the monotone function $f_0(E, L^2)$ (the results obtained by the authors quoted above are for isotropic systems, but there is no problem to generalize them to the anisotropic case, as we do here). Formally, $W_2$ is easily shown to satisfy the relation

$$W_2[-\{f_0, \overline{g_1}\}] = H^{(2)}[\overline{g_1}] \tag{61}$$

for any function $g_1$. The admissible $\mathcal{Q}_2$ is taken to be the set of all the $\overline{f_1}$ satisfying the constraints

$$\int \overline{f_1} \, \mathrm{d}\xi = 0 \quad \text{and} \quad \int \overline{f_1} C'(f_0) \, \mathrm{d}\xi = 0 , \tag{62}$$

which imply the conservation of mass and $H_C$-function to first order. The constraints which are retained from VP are thus quite weak, and, as a consequence, it has not been possible up to now to derive in this framework stability results as complete as those furnished by the energy and the symplectic methods.

c) The functional used by (Sygnet et al. 1984) appears to be $W_2[\overline{f_1}]$, as in the previous method, but the set of admissible functions $\overline{f_1}$ is not clearly specified. The authors, however, effect without further justifications the change of dependant variable $\overline{f_1} \to \mu$ defined by

$$\overline{f_1}(r, E, L^2) = \left[ (rv_r) f_{0E} \frac{\partial(rv_r \mu)}{\partial r} \right] (r, E, L^2) \tag{63}$$

[their Eq. (B3), written with our notations], where use is made of $(r, E, L^2)$ as independant variables [actually, this is a proper choice only in the region $\{v_r > 0\}$ or in the region $\{v_r < 0\}$]. But such a change can be valid only for particular perturbations. First, it is easy to see that any $\overline{f_1}$ of the form Eq. (63) is symplectic, being generated by $\overline{g_1}(r, E, L^2) := (rv_r \mu)(r, E, L^2)$. Secondly, the new function $\mu$ introduced by this relation can be regular only if $\overline{g_1} = 0$ for $rv_r = 0$. In our work, on the contrary, these conditions are explicit part of a coherent formalism: The symplectic nature of the perturbations is a basic element of the framework, while the vanishing of $\overline{g_1}$ for $rv_r = 0$ is automatically ensured by restricting our attention to the antisymmetric part $\overline{g_1}^-$ of $\overline{g_1}$, the part of the functional related to $\overline{g_1}^+$ turning out to be trivially positive.

## 5   STABILITY WITH RESPECT TO ASPHERICAL PERTURBATIONS

### 5.1   A further decomposition

Using the relation

$$\delta f_1 = -\{f_0, \delta g_1\} = -f_{0E}\{E, \delta g_1\} - f_{0L^2}\{L^2, g_1\} , \tag{64}$$

we can write the quantity $H^{(2)}[\delta g_1]$ as a sum of two terms:

$$H^{(2)}[\delta g_1] = H_1^{(2)}[\delta g_1] + H_2^{(2)}[\delta g_1] , \tag{65}$$

with

$$H_1^{(2)}[\delta g_1] := \frac{1}{2} \int \frac{|\{f_0, \delta g_1\}|^2}{-f_{0E}} \, d\xi - \frac{1}{8\pi G} \int |\nabla_x \delta \Phi_1|^2 \, d\mathbf{x} \,, \tag{66}$$

$$H_2^{(2)}[\delta g_1] := -\frac{1}{2} \int \frac{f_{0L^2}}{-f_{0E}} \{L^2, \delta g_1^*\}\{f_0, \delta g_1\} \, d\xi \,. \tag{67}$$

### 5.2 Sign of $H_1^{(2)}[\delta g_1]$

We now set

$$\delta f_1 = -\{f_0, \delta g_1\} =: -f_{0E}\delta\Phi_1 + \delta\tilde{f}_1 \,, \tag{68}$$

i.e., we use the standard trick [introduced in the gravitational context by (Lynden-Bell 1966)] which consists to single out the "quasi-static part" in the variation $\delta f_1$ of the distribution function. Then, taking into account the relation

$$\int \delta f_1 \delta\Phi_1 \, d\xi = -\frac{1}{4\pi G} \int |\nabla_x \delta\Phi_1|^2 \, d\mathbf{x}, \tag{69}$$

we can write

$$\begin{aligned}
H_1^{(2)}[\delta g_1] &= \frac{1}{2} \int \frac{(\delta f_1)^2}{-f_{0E}} \, d\xi - \frac{1}{8\pi G} \int |\nabla_x \delta\Phi_1|^2 \, d\mathbf{x} \\
&= \frac{1}{2} \int \left[ \frac{(\delta\tilde{f}_1)^2}{-f_{0E}} - f_{0E}(\delta\Phi_1)^2 + (\delta f_1 + f_{0E}\delta\Phi_1)\delta\Phi_1 \right] d\xi - \frac{1}{8\pi G} \int |\nabla_x \delta\Phi_1|^2 \, d\mathbf{x} \\
&= \frac{1}{8\pi G} \int \left[ |\nabla_x \delta\Phi_1|^2 - 4\pi G \left( \int (-f_{0E}) \, d\mathbf{v} \right) |\delta\Phi_1|^2 \right] d\mathbf{x} + \frac{1}{2} \int \frac{(\delta\tilde{f}_1)^2}{-f_{0E}} \, d\xi \,.
\end{aligned} \tag{70}$$

Following (Aly & Perez 1992), we now set

$$\delta\Phi_1 =: \Psi_0' w. \tag{71}$$

Then

$$\begin{aligned}
H_1^{(2)}[\delta g_1] &= \frac{1}{2} \int \frac{(\delta\tilde{f}_1)^2}{-f_{0E}} \, d\xi + \frac{1}{8\pi G} \int \left[ \Psi_0'^2 \left( |\nabla_x w|^2 - 4\pi G(-f_{0E})|w|^2 \right) + |w|^2 (\nabla_x \Psi_0')^2 + \Psi_0' \nabla_x \Psi_0' . \nabla_x |w|^2 \right] d\mathbf{x} \\
&= \frac{1}{2} \int \frac{(\delta\tilde{f}_1)^2}{-f_{0E}} \, d\xi + \frac{1}{8\pi G} \int \left[ (\Psi_0')^2 \left( |\nabla_x w|^2 - 4\pi G \left( \int (-f_{0E}) \, d\mathbf{v} \right) |w|^2 \right) - |w|^2 \Psi_0' \nabla_x^2 \Psi_0' \right] d\mathbf{x} \,.
\end{aligned} \tag{72}$$

The quantity $\nabla_x^2 \Psi_0'$ can be computed by differentiating Eq. (14) with respect to $r$, which gives

$$\nabla_x^2 \Psi_0' = 4\pi G(\rho_0' + \rho_e') + 2\frac{\Psi_0'}{r^2} = 4\pi G \left[ \int \left( \Psi_0' f_{0E} + 2rv_t^2 f_{0L^2} \right) d\mathbf{v} + \rho_e' \right] + 2\frac{\Psi_0'}{r^2} \,. \tag{73}$$

On the other hand, $w$ has zero average value on any spherical surface of center $O$, and then, by the so-called Wirtinger inequality (Aly & Perez 1992), we have for any value of $r$

$$\int \left( |\nabla_{xs} w|^2 - \frac{2}{r^2}|w|^2 \right) d\Omega \geq 0 \,, \tag{74}$$

with equality holding if and only if $w$ is of the form $w(r, \theta, \phi) = w_0(r)\hat{\mathbf{x}}.\mathbf{a}$ for some constant vector $\mathbf{a}$ [in Eq. (74), we have set $\nabla_{xs} := \nabla_x - \hat{\mathbf{r}}\partial/\partial r$ and $d\Omega := \sin\theta d\theta d\phi$] .

Using these results in Eq. (72), we obtain eventually

$$\begin{aligned}
H_1^{(2)}[\delta g_1] &= \frac{1}{2} \int \frac{(\delta\tilde{f}_1)^2}{-f_{0E}} \, d\xi + \frac{1}{8\pi G} \int \left[ (\Psi_0')^2 \left( \left|\frac{\partial w}{\partial r}\right|^2 + |\nabla_{xs} w|^2 - 2\frac{|w|^2}{r^2} \right) + 4\pi G \Psi_0' \left( (-\rho_e') + 2r \int v_t^2(-f_{0L^2}) \, d\mathbf{v} \right) \right] d\mathbf{x} \\
&\geq \frac{1}{2} \int \frac{(\delta\tilde{f}_1)^2}{-f_{0E}} \, d\xi + \frac{1}{8\pi G} \int \left[ (\Psi_0')^2 \left|\frac{\partial w}{\partial r}\right|^2 + 4\pi G \Psi_0' \left( (-\rho_e') + 2r \int v_t^2(-f_{0L^2}) \, d\mathbf{v} \right) |w|^2 \right] d\mathbf{x} \,.
\end{aligned} \tag{75}$$

Therefore, $H_1^{(2)}[\delta g_1] \geq 0$ for all $g_1$ if

$$\rho_e' \leq 0 \,, \tag{76}$$

and

$$f_{0L^2} \leq 0 \,. \tag{77}$$

Let us now assume that both conditions are satisfied, and that $H_1^{(2)}[\delta g_1] = 0$. Then: (i) $\delta\tilde{f}_1 = 0$; (ii) $\partial w/\partial r = 0$ and Wirtinger's inequality reduces to an equality, whence $w = \hat{\mathbf{x}}.\mathbf{a}$. (iii) If either $\rho_e' \not\equiv 0$ or $f_{0L^2} \not\equiv 0$, we must also have $w = 0$ for some values of $r$. Then $w \equiv 0$ everywhere, and $\delta f_1 \equiv 0$. (iv) If $\rho_e' \equiv 0$ (which implies $\rho_e \equiv 0$ and $\Phi_e \equiv 0$) and $f_{0L^2} \equiv 0$, the previous results and the definition of $w$ imply $\delta f_1 = -f_{0E}\Psi'_0 \hat{\mathbf{x}}.\mathbf{a} = -\mathbf{a}.\nabla_x f_0$, which means that the perturbation just results from a rigid translation of the system of a vector $\mathbf{a}$. But there is no such perturbations with $\mathbf{a} \neq 0$ generated by the elements of $\mathcal{G}$ (see §3), and then we have in that case too $w \equiv 0$ and $\delta f_1 \equiv 0$. We have thus proven eventually that, under conditions (76) and (77),

$$H_1^{(2)}[\delta g_1] > 0 \quad \text{if} \quad \delta f_1 \not\equiv 0 . \tag{78}$$

### 5.3  Sign of $H_2^{(2)}[\delta g_1]$

Clearly, the term $H_2^{(2)}[\delta g_1]$ can be of either sign, depending on the choice of the generator. We even know from the general theorem of Paper I quoted in §3.2, that, by an appropriate choice of $g_1$, it can be made sufficiently negative to overcome the positivity of $H^{(2)}[\overline{g_1}] + H_1^{(2)}[\delta g_1]$ which holds under conditions (76) and (77). Then we could only try to classify the perturbations according to the sign of that quantity.

Even that more modest goal appears to be difficult to reach, and we shall content ourselves here with the simple following remark: There is a particular class of perturbations, defined by the condition

$$f_{0L^2}\{L^2, g_1\} = f_{0L^2}\{L^2, \delta g_1\} = 0 , \tag{79}$$

for which $H_2^{(2)}[\delta g_1]$ vanishes [note the use in Eq. (79) of the obvious relation $\{L^2, \overline{g_1}\} = 0$]. We feel that these elements of $\mathcal{G}$ – which thus commute with $L^2$ in the region of the phase space where $f_{0L^2} \neq 0$ – play some role in the dynamics of the system, and then they deserve a particular name: We shall call them "preserving generators" – the associated perturbations being the "preserving perturbations" –, and denote as $\mathcal{P}$ their set.

Clearly, $\mathcal{P}$ is a linear subspace of $\mathcal{G}$, and the general form of its elements [which do automatically satisfy condition (26)] can be found as follows. Consider the part $\Omega_1$ of $\Omega_0$ where $f_{0L^2} \neq 0$. We first note that there are five obvious solutions to Eq. (79) in $\Omega_1$, namely: $r$, $p_r$, $L_x = -\cot\theta\cos\phi p_\phi - \sin\phi p_\theta$, $L_y = -\cot\theta\sin\phi p_\phi + \cos\phi p_\theta$ and $L_z = p_\phi$. Moreover, they are linearly independant, which can be most easily checked by considering the Jacobian determinant $D(r, p_r, L_x, L_y, L_z)/D(r, \theta, \phi, p_r, p_\theta, p_\phi)$, which turns out to be of rank 5. As the general solution to Eq. (79) in $\Omega_1$ must depend on at most five integrals, we can thus conclude that any preserving generator is of the form

$$g_1(r, \theta, \phi, p_r, p_\theta, p_\phi) = g_1(r, p_r, p_\phi, -\cot\theta\cos\phi p_\phi - \sin\phi p_\theta, -\cot\theta\sin\phi p_\phi + \cos\phi p_\theta) , \tag{80}$$

in $\Omega_1$, and arbitrary elsewhere. Of course, all the spherically symmetric $g_1$ belong to $\mathcal{P}$. But this set contains many other elements, as the spherically symmetric elements depend on three variables [$g_1 = \overline{g_1} = g_1(r, p_r, L_x^2 + L_y^2 + L_z^2)$], while the general preserving ones depend on five variables. Unfortunately, $\mathcal{P}$ is not "closed under evolution" (see §3.2). Indeed, it results at once from the equation of evolution (25) that condition (79) is time-invariant only if the potential $\Phi_1$ stays spherically symmetric, which can be the case only for the particular spherically symmetric perturbations.

We can thus eventually summarize the results of §5.2-5.3 in the form of

*Theorem 2:* If the conditions (18), (76) and (77) are satisfied, then we have $H^{(2)}[\delta g_1] > 0$ for all the nontrivial aspherical generators of $\mathcal{G}$ satisfying $f_{0L^2}\{L^2, \delta g_1\} = 0$.

### 5.4  Consequences for stability

From Theorem 2 above and the general statements of §3.2, we can first conclude at once that an isotropic system is stable with respect to any aspherical perturbations – and then is stable with respect to any perturbation, because of the result of §4.2, – if condition (76) is satisfied, i.e., if the external mass density $\rho_e$ is a nonincreasing function of $r$ (which is always the case in the applications).

For anisotropic systems satisfying (76) and (77), our Theorem 2 shows that any nontrivial aspherical perturbation generated by a preserving $\delta g_1$ has positive energy – the same result holding true for any preserving $g_1$ because of theorem 1. This completely new result is unfortunately more difficult to interpret. Strictly speaking, neither of the general stability criteria established thus far do apply here, as the preserving character of a perturbation is conserved in time only for spherically symmetric perturbations ($\{f_0, g_1\} = 0$). However, we feel that our result has some relevance for the stability problem. Numerical simulations giving some support to this idea are reported in (Perez et al. 1995).

### 5.5  Comparison with previous works

For isotropic systems, the stability result above has already been derived by the thermodynamic method and the energy principle ((Antonov 1962), (Lynden-Bell & Sannit 1969), (Binney & Tremaine 1987), where it is referred to as the Antonov-Lebowitz theorem) in the case where there is no external potential (i.e., $\Phi_e \equiv 0$). The new proof we have presented here, however, is more simple and transparent, and it is self-contained (e.g., it does not use results on the stability of gaseous stars). Moreover, it does not appeal to some unproven assumptions on the completeness of the modes of some operator (assumptions which may certainly be proven, but whith much efforts). Our work thus answers an important question addressed by (Kandrup & Sygnet 1985) about the possibility of a stability proof not using some peculiar completeness theorem, which certainly justified to present it in details.

Concerning the aspherical stability of an anisotropic system, there seems to be only one published result based on the consideration of an energy functional. It is due to (Gillon et al. 1976), and states that any spherically symmetric steady state satisfying $\Phi_e \equiv 0$ and conditions (18) and (77) is stable with respect to any aspherical perturbations. There has been some doubts repeatedly casted in the literature on the validity of this result (e.g., (Binney & Tremaine 1987)), but no explicit rebuttal (or approval!) has been yet presented. Owing to its possible importance, we have thus conducted a detailed analysis of Gillon et al.'s complicated proof, in which a spherical system is approximated by a multiple water-bag model. Unfortunately, we have found two serious flaws, and our conclusion does confirm the fear of the skeptics:

a) There is an inconsistency in the calculations, as the conservation law for the basic quantity $A$ [their Eq. (49)] is obtained by using both the nonlinear and the linearized VP, without distinguishing carefully the order of the various terms. When the derivation is made correctly, it does not seem to lead to an interesting result.

b) If we forget about this inconsistency and admit for a while the validity of their Eq. (49), then we must note that the "conserved" quantity $A$ [their Eq. (52)] takes a positive value $A_{eq} > 0$ at equilibrium [just make $a_{\pm k} = 0$ in their Eq. (52), where $a_{\pm k}$ are the functions describing an arbitrary perturbation]. Therefore, to be a Lyapunov function, $A$ should satisfy the condition $A \geq A_{eq}$ for any perturbation rather than the stated one, $A \geq 0$. But it is easy to check that the former condition is certainly not fulfilled, as $A$ can be made larger or smaller than $A_{eq}$ by choosing adequately $a_{\pm k}$. Then, even under our optimistic assumption, no conclusion could be actually drawn from the sign of $A$.

## 6 EXTENSION TO OTHER SYSTEMS

### 6.1 Rotating spherical systems

Consider a spherically symmetric steady state characterized by the function $f_s(E, L^2)$ and the potential $\Psi_0$, and introduce an arbitrary function $f_a(E, L^2, L_z)$ such that

$$f_0(E, L^2, L_z) := f_s(E, L^2) + f_a(E, L^2, L_z) \geq 0, \tag{81}$$

and

$$f_a(E, L^2, -L_z) = -f_a(E, L^2, L_z). \tag{82}$$

As noted long ago by (Lynden-Bell 1960), $f_0$ is also a steady state distribution function associated to the potential $\Psi_0$ (the mass density corresponding to $f_0$ is the same as the one corresponding to $f_s$), and it describes a system which is rotating about the $z$-axis. We now discuss whether the stability results derived above can be extended to these more general distribution functions.

Proceeding as in §3.3, we obtain after some algebra

$$\begin{aligned} H^{(2)}[g_1] &= -\frac{1}{2} \int \{f_s, \overline{g_1}\}\{E, \overline{g_1}^*\} \, d\xi - \frac{1}{2} \int \{f_0, \delta g_1\}\{E, \delta g_1^*\} \, d\xi - \int \{f_a, \overline{g_1}\}\{E, \delta g_1^*\} \, d\xi \\ &\quad - \frac{1}{8\pi G} \int |\nabla \overline{\Phi_1}|^2 \, d\mathbf{x} - \frac{1}{8\pi G} \int |\nabla \delta \Phi_1|^2 \, d\mathbf{x}, \end{aligned} \tag{83}$$

with

$$\overline{\Phi_1} = K[-\{f_s, \overline{g_1}\} - \overline{\{f_a, \delta g_1\}}], \tag{84}$$

$$\delta\Phi_1 = K[-\{f_s, \delta g_1\} - (\{f_a, \delta g_1\} - \overline{\{f_a, \delta g_1\}})]. \tag{85}$$

Because of the presence of the nonspherically symmetric term $f_a$ in $f_0$, it is then no longer possible to separate $H^{(2)}[g_1]$ into two terms depending respectively on $\overline{g_1}$ and $\delta g_1$, and the analysis becomes rather intricate.

However, there are two simple – but new – results which can be extracted from the expressions above when $\partial f_s/\partial E < 0$. Consider first a purely spherically symmetric perturbation generated by $\overline{g_1}$. Then

$$H^{(2)}[g_1] = H^{(2)}[\overline{g_1}] = -\frac{1}{2} \int \{f_s, \overline{g_1}\}\{E, \overline{g_1}^*\} \, d\xi - \frac{1}{8\pi G} \int |\nabla \overline{\Phi_1}|^2 \, d\mathbf{x}, \tag{86}$$

$$\overline{\Phi_1} = K[-\{f_s, \overline{g_1}\}], \tag{87}$$

i.e., we recover the expressions valid for the spherical state $f_s$. It thus results at once from §4 that $H^{(2)}[\overline{g_1}] > 0$ for any nontrivial $\overline{g_1}$, and our rotating state turns out to be stable with respect to all the spherically symmetric perturbations (their character being preserved in time).

Consider now a system with $f_s = f_s(E)$ which is submitted to some axisymmetric perturbation ($\partial \delta g_1/\partial \phi = 0$) satisfying $\delta g_1(-v_\phi) = \delta g_1(v_\phi)$ (a property also conserved in time). Then stability is obtained in that case too. As easily checked indeed, the cross term in Eq. (83) [containing both $\overline{g_1}$ and $\delta g_1$] and the perturbed potential associated to $f_a$ vanish, and the results of §4-5 imply $H^{(2)}[g_1] > 0$.

### 6.2 Systems with a nondecreasing distribution function

The stability of a spherical system when condition (18) is not satisfied has been considered numerically by (Henon 1973). From an analytical point of view, it appears to be a quite difficult problem, and here, we shall content ourselves to make two points.

a) Consider the case where $f_{0E} > 0$ in the region $\Omega_0$ where $f_0 > 0$. Of course, this implies that $f_0$ is discontinuous on the boundary $\partial \Omega_0 = \{\xi \,|\, E(\xi) = E_0 \leq 0\}$, and there is a violation of our smoothness assumption. Consider a spherical perturbation generated by $\overline{g_1}$, and assume that $\{E, \overline{g_1}\}$ vanishes on $\partial\Omega_0$. It is easy to check that $H^{(2)}[\overline{g_1}]$ is still given by Eq. (39) [the boundary term related to the presence of the delta-function $\delta(E - E_0)$ in the derivative $f_{0E}$ disappearing], and that $H^{(2)}[\overline{g_1}] < 0$ for any nontrivial $\overline{g_1}$. Then the system appears to be stable with respect to these particular perturbations (§3.2.a) – a conclusion also obtained by (Kandrup & Sygnet 1985) by means of their energy principle.

b) Consider the case where the system is isotropic $[f_0 = f_0(E)]$ and $f_{0E}$ may have a changing sign in $\Omega_0$, and let us assume that the system admits a neutral aspherical mode generated by $\delta g_1$. Then we have $\delta \tilde{f}_1 = 0$ and $H^{(2)}[\delta g_1] = 0$ (Paper I), which implies at once, by using the arguments at the end of §5.1, that $\delta f_1 = 0$. We have thus proven the following new result: Whichever be the sign of $f_{0E}$, an isotropic spherically symmetric system cannot admit an aspherical neutral mode.

## 7  CONCLUSION

In this paper, we have considered a self-gravitating collisionless system, in which the particles are also submitted to an external potential $\Phi_e$. We have first presented some new considerations on the relations between the spherical symmetry of a steady state and the form of its distribution function $f_0$. In particular, we have argued that, under quite general assumptions, a system with an $f_0$ depending only on $E$ and $L^2$ is necessarily spherically symmetric.

We have thus reinvestigated from an analytical point of view the problem of the linear stability of this system in the consistent framework of the symplectic method. By effecting a systematic study of the sign of the energy of an arbitrary symplectic perturbation, we have reached in particular the following conclusions, valid when $f_{0E} < 0$:

a) Any steady state is stable with respect to all the spherically symmetric perturbations.

b) An isotropic system $[f_0 = f_0(E)]$ is also stable with respect to all the aspherical perturbations, at least if the external mass density $\rho_e$ is a nonincreasing function of the radial coordinate $r$. Of course, this result and the one in (a) above are well known in the case where $\Phi_e \equiv 0$. But even if we restrict our attention to this particular case, our proofs are much simpler and complete than all the previous ones, bringing about some new important elements which were called for by many authors. In particular, our proof of the second quoted result shows more clearly the origin of the stability, and it is the first which does not appeal to some unproven assumption on the completeness of a set of modes.

c) For an anisotropic system with $f_{0L^2} \leq 0$ and $\rho'_e \leq 0$, any nontrivial "preserving perturbation" – i.e., one having a generator commuting with $L^2$, which of course does not mean that it is spherically symmetric – has a positive energy. The exact meaning of this result is still not completely clear, and we have tried to elucidate it by effecting numerical investigations. The results obtained thus far [in particular for the case where the distribution function is of the Ossipkov-Meritt type (see, e.g., (Binney & Tremaine 1987)) and there is no external potential] are reported in the companion paper (Perez et al. 1995). More general situations, currently under intensive study, will be analyzed in forthcoming papers.

Finally:

a) We have analyzed in details a challenged result of (Gillon et al. 1976), according to which stability with respect to any perturbation holds if $f_{0L^2} \leq 0$, and found the proof to contain some flaws, that we have clearly explicited.

b) We have shown that a spherical rotating system is stable with respect to all the spherically symmetric perturbations, and [when $f_s = f_s(E)$] to all the axisymmetric even ones.

c) We have proven that an isotropic system with a derivative $f_{0E}$ which may change its sign, cannot admit a nontrivial neutral mode (i.e., one differing from a mere global translation of the system).